\documentclass[showpacs,aps,floatfix,prb,reprint,superscriptaddress]{revtex4-1}
\usepackage{graphicx}
\usepackage{xfrac}
\usepackage{amssymb}
\usepackage{tikz}
\usepackage{amsfonts} 
\usepackage{amsmath} 
\usepackage[utf8]{inputenc} 
\usepackage{mathtools}
\usepackage{xcolor,colortbl}
\usepackage{bm} 
\usepackage{array}
\usepackage{color}  
\usepackage{relsize} 
\newcommand{\V}[1]{\vec{#1}}

\newcommand{\bk}{\mathbf{k}}

\newcommand{\bra}[1]{\langle #1 |}
\newcommand{\ket}[1]{| #1 \rangle}

 \usepackage[english]{babel}
 \usepackage{amssymb,amsmath}
 \usepackage{color}
 \usepackage{pstricks}
 \usepackage{bbold}
 \usepackage{graphicx}
 \usepackage{color}
 \usepackage{verbatim}
 \usepackage{fancyvrb}
 \usepackage{alltt}
 \usepackage{longtable}
 \usepackage{amsmath}

\begin{document}

\title{Accurate \textit{ab initio} tight-binding Hamiltonians: effective tools for electronic transport and optical spectroscopy from first principles}

\author{Pino D'Amico} 
\affiliation{Dipartimento di Fisica, Informatica e Matematica, Universit\'a di Modena and Reggio Emilia, Via Campi 213/a, 41125 Modena, Italy}
\affiliation{CNR-NANO Research Center S3, Via Campi 213/a, 41125 Modena, Italy}

\author{Luis Agapito} 
\affiliation{Department of Physics and Department of Chemistry, University of North Texas, Denton, TX 76203, USA}
\affiliation{Center for Materials Genomics, Duke University, Durham, NC 27708, USA}

\author{Alessandra Catellani}
\affiliation{CNR-NANO Research Center S3, Via Campi 213/a, 41125 Modena, Italy}

\author{Alice Ruini} 
\affiliation{Dipartimento di Fisica, Informatica e Matematica, Universit\'a di Modena and Reggio Emilia, Via Campi 213/a, 41125 Modena, Italy}
\affiliation{CNR-NANO Research Center S3, Via Campi 213/a, 41125 Modena, Italy}

\author{Stefano Curtarolo}
\affiliation{Center for Materials Genomics, Duke University, Durham, NC 27708, USA}
\affiliation{Materials Science, Electrical Engineering, Physics and Chemistry, Duke University, Durham, NC 27708, USA}

\author{Marco Fornari}
\affiliation{Department of Physics, Central Michigan University and Science of Advanced Materials Program, Mt. Pleasant, MI 48859}
\affiliation{Center for Materials Genomics, Duke University, Durham, NC 27708, USA}

\author{Marco \surname{Buongiorno Nardelli}} 
\email{Email: mbn@unt.edu}
\affiliation{Department of Physics, University of North Texas, Denton, TX 76203, USA}
\affiliation{Center for Materials Genomics, Duke University, Durham, NC 27708, USA}

\author{Arrigo Calzolari}
\email {Email: arrigo.calzolari@nano.cnr.it}
\affiliation{CNR-NANO Research Center S3, Via Campi 213/a, 41125 Modena, Italy}
\affiliation{Department of Physics, University of North Texas, Denton, TX 76203, USA}
\affiliation{Center for Materials Genomics, Duke University, Durham, NC 27708, USA}

\date{\today}

\begin{abstract}
The calculations of electronic transport coefficients and optical properties require a very dense interpolation of the electronic band structure in reciprocal space that is computationally expensive and may have issues with band crossing and degeneracies. Capitalizing on a recently developed pseudo-atomic orbital projection technique, we exploit the exact tight-binding representation of the first principles electronic structure for the purposes of (1) providing an efficient strategy to explore the full band structure $E_n({\bf k})$, (2) computing the momentum operator differentiating directly the Hamiltonian, and (3) calculating the imaginary part of the dielectric function.  This enables us to determine the Boltzmann transport coefficients and the optical properties within the independent particle approximation. In addition, the local nature of the tight-binding representation facilitates the calculation of the ballistic transport within the Landauer theory for systems with hundreds of atoms. In order to validate our approach we study the multi-valley band structure of CoSb$_3$ and a large core-shell nanowire using the ACBN0 functional. In CoSb$_3$ we point the many band minima contributing to the electronic transport that enhance the thermoelectric properties; for the core-shell nanowire we identify possible mechanisms for photo-current generation and justify the presence of protected transport channels in the wire.
\end{abstract}

\maketitle

\section{introduction} 

The ability to efficiently generate and manage a combination of theoretical and experimental data is the foundation for data driven discovery of new materials and functions as well as methods to control manufacturing processes.\cite{MRS:7963345,curtarolo:nmat_review} 
This formidable task requires a continuous feedback loop where descriptors\cite{curtarolo:nmat_review}  of the functional properties are calculated for an enormous number of materials configurations, integrated in the databases,\cite{Curtarolo2012227,Curtarolo2012218,Jain2013,Pizzi2016218} compared with the available experiments, and exploited in the prediction cycle. In this paper we focus on a broad class of descriptors derived from the electronic structure calculations in order to provide easier integration with the experimental data. We introduce tight-binding methodologies for the calculation of electronic transport properties and the simulation of optical spectroscopies in the broadest energy range and with excellent accuracy as well as high computational efficiency. 

The  prerequisite for the simulation of both electron transport and optical properties is the accurate evaluation of the electronic structure of the system that is obtained by a fully self-consistent quantum-mechanical calculation either within density functional theory (DFT) or other first principles approaches. The electronic structure of solids is often described with Fourier basis functions that account naturally for the the periodicity and whose completeness is easily improvable up to any desirable accuracy. The delocalized character of plane waves, however, is not appropriate for the description of highly localized electronic systems unless a very large number of basis functions is used. The development of minimal-space solutions such as atomic orbital (AO) Bloch sums, which are capable of capturing with satisfactory accuracy the properties of solids and molecules on finite Hilbert spaces, has been central to methodological developments in quantum chemistry and solid state physics for many decades. 

Atomic orbital basis sets provide the foundation for methods, such as tight-binding (TB), which combine
an intuitive physical representation of the interactions, low computational cost, and interesting alternatives for the study of the electronic structure of  molecules and solids.\cite{Huckel:1931ft, Jones:1934go,Harrison:1989}
The TB Hamiltonian matrix is the central quantity that  provides a compact real space representation of the many-body interactions, the accuracy of the electronic structure relies on the quality of such matrix.
Albeit computationally inexpensive and very intuitive for simple compounds, semiempirical TB implementations often fail in the prediction of electronic structure of complex materials and, in most cases, lack of predictive value when dealing with structural and chemical modifications as well as charge rearrangements.
In recent years, the reliability of the TB models has been largely improved with the introduction of \textit{ab initio} TB Hamiltonians derived from fully self-consistent quantum-mechanical calculations through a mapping into a much smaller space spanned by a set of atomic or atomic-like orbitals.
This procedure combine the accuracy and the predictive value of first principles approaches with the low computational cost of TB techniques.  Furthermore,  it is particularly useful for the evaluation of properties such as the electron conductivity and the optical absorption spectra that require a precise and ultrafine reciprocal space integration,  typically very computationally expensive. 

We recently developed a straightforward, non-iterative projection scheme that can exactly represent the first principles electronic structure of a periodic system on a finite AO-like basis.\cite{Agapito_2013_projectionsPRB, Agapito:2016jl, Agapito:2016en} By filtering the projections of Bloch states with high-kinetic-energy components and tuning the richness of the finite Hilbert space, we construct fully first principles TB Hamiltonians where the number of exactly reproduced bands with respect to the original DFT calculation can be selectively increased at a negligible computational cost. This provides cost effective solutions to design efficient algorithms for electronic structure simulations of realistic material systems and massive high-throughput investigations. 
Our technique does not seek construction of (heavily customized, localized) basis functions. Its value resides on allowing non-iterative reproduction of a large number of energy bands using standard quantum-chemistry basis sets or the pseudo-atomic orbitals (PAO) of a standard pseudopotential calculation. Practically, the present methodology completely supersedes the need for engineered basis functions such as maximally localized Wannier functions\cite{Marzari:1997p231} or muffin-tin orbitals of arbitrary order (NMTO)\cite{Andersen:2000fc}  in the context of the evaluation of transport \cite{Mostofi:2014hw,Calzolari:2004p39,Pizzi:2014ee} or optical properties.
Moreover, the knowledge of a localized orbital representation that maps seamlessly the electronic structure onto a localized AO basis set that is eventually fitted to a Gaussian basis set, opens the way to the fast (analytical) computation of two-electron integrals for solid-state applications and it is at the core of the development of the accurate and efficient ACBN0 functional.\cite{Agapito:2015iz} 

The paper is organized as follows: in Sec. \ref{method} we will discuss the theoretical background and the practical implementation of the procedure with representative test cases. In Sec. \ref{results} we study, with our methodology, two significant materials problems in order to show the importance of fine reciprocal space sampling and the computational efficiency of the PAO projection to deal with very large systems. We chose to study the thermolectric properties of CoSb$_3$ and the transport and optical properties of core-shell nanowire of ZnO and ZnS. Finally in Sec. \ref{conclusions} we outline the conclusion of the present work.

\section{Methodology} \label{method}

\subsection{TB representation from PAO projections} 

Accurate TB Hamiltonian matrices can be built from the direct projection of the Kohn-Sham (KS) Bloch states $\ket{\psi_{n\bk}}$ onto a chosen  basis set of fixed localized functions, as we discussed extensively in Ref.~\onlinecite{Agapito_2013_projectionsPRB,Agapito:2016jl,Agapito:2016en}.
There, we have shown that the real space Hamiltonians $ \hat{H}\left({\bf r}_{\alpha}\right)$ can be directly calculated using atomic orbitals or PAOs from the pseudopotential of any given element. The key in this procedure is in the mapping of the {\it ab initio} electronic structure (solved on a well converged and large plane waves basis set) into a model that precisely reproduces a selected number of bands of interest.\cite{Agapito_2013_projectionsPRB, Agapito:2016jl} 


The crucial quantities that measure the accuracy of the basis set are the projectabilities $p_{n\bk}=\bra{\psi_{n\bk}} \hat{P} \ket{\psi_{n\bk}} \ge 0$ ($\hat{P}$ is the operator that projects onto the space of the PAO basis set, as defined in 
Ref.~\onlinecite{Agapito:2016jl}
) which indicate  the representability of a Bloch state $\ket{\psi_{n\bk}}$ on the chosen PAO set.
Maximum projectability, $p_{n\bk}= 1$, indicates that the particular Bloch state can be perfectly represented in the chosen PAO set; contrarily, $p_{n\bk} \approx 0$ indicates that the PAO set is insufficient and should be augmented. 
Once the Bloch states with good projectabilities have been identified, the TB Hamiltonian is constructed as: 
\begin{equation}
\hat{H}(\bk) = AEA^\dagger + \kappa \left( I-A \left( A^{\dagger}A \right)^{-1}A^\dagger \right)
\label{eq:Hk}
\end{equation}
where $E$ is the diagonal matrix of KS eigenenergies and $A$ is the matrix of coefficients obtained from projecting the Bloch wavefunctions onto the PAO set (See Ref.~\onlinecite{Agapito:2016jl}.) Since the filtering procedure introduces a null space, the parameter $\kappa$ is used to shift all the unphysical solutions outside a given energy range of interest.

The procedure provides an accurate real space representation of the {\it ab initio} Hamiltonian $\hat{H}\left({\bf r}_{\alpha}\right)$ as a TB matrix of very small dimension, a crucial advantage for the accurate calculation of any physical properties that requires the precise integration in the reciprocal space. 




By exploiting the PAO projection scheme described above we can easily Fourier transform the TB real space representation, interpolate to arbitrary precision, and perform derivatives in reciprocal space. For example, 
the expectation value of the momentum operator, which is the main quantity in the definition of both and transport and optical descriptors described below, is computed as :    

\begin{eqnarray}\label{momentum}
{\bf p}_{nm} ({\bf k})&=&  \left < \psi_n ({\bf k}) \right |\hat{p}\left | \psi_m ({\bf k}) \right > = \\ \nonumber
&=&  \left < \psi_n ({\bf k}) \right |\frac{m_0}{\hbar} \V{\nabla}_{\bf k} \hat{H}({\bf k})\left | \psi_m ({\bf k}) \right > 
\end{eqnarray}

with 

\begin{equation} \label{Hamiltonian.gradient}
\V{\nabla}_{{\bf k}} \hat{H}({\bf k}) = 
\sum_{\alpha} i{\bf r}_{\alpha} \exp\left({i{\bf k}\cdot {\bf r}_{\alpha}}\right) \hat{H}\left({\bf r}_{\alpha}\right).
\end{equation}

$\hat{H}\left({\bf r}_{\alpha}\right)$ being the real space TB matrix and  $\left | \psi_n ({\bf k}) \right >$ the eigenstates of the Hamiltonian $\hat{H}({\bf k})$.

\subsection{Boltzmann transport} \label{boltzmann}

Within the semiclassical theory, the electrical conductivity can be evaluated by solving the Boltzmann equation that describes 
the evolution of the distribution function $f$ of an electron gas under external electric field and in presence of scattering mechanisms.\cite{Parravicini:2000ud,Singh2001125,Madsen200667} 
In the so-called scattering-time approximation,  the conductivity tensor $\sigma_{ij}$ can be expressed as an integral over the Brillouin Zone (BZ):
\begin{equation} \label{boltzmann.conductivity.equation.generalized}
\sigma_{ij}= \frac{e^2}{4\pi^3} \int_{BZ} \tau \sum_n  v_n^i({\bf k})v_n^j({\bf k}) \left( -\frac{\partial f_0}{\partial \epsilon}\right)d {\bf k},
\end{equation}
where $\tau$ is the relaxation time, $v_n^i({\bf k})$ is the {\em i}-th component of the  electron velocity (${\bf v}_n$) corresponding to the  {\em n}-th band for each {\bf k}-point in the BZ,  $f_0$ is the equilibrium distribution function, and $\epsilon$ is the electron energy.

\begin{figure}[h!]
\begin{center}
    \includegraphics[width=0.95\columnwidth]{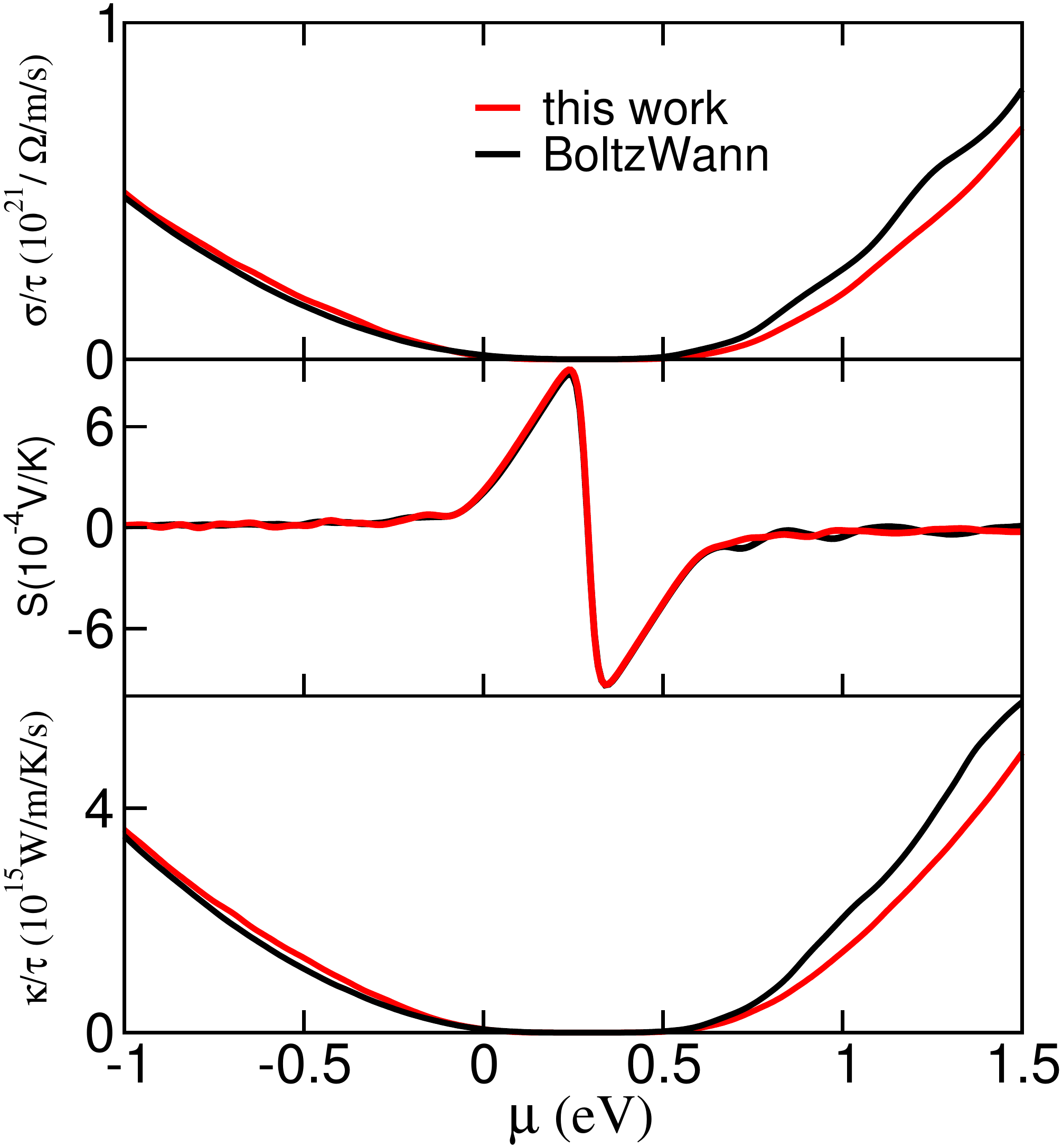}
    \vspace{5mm}
    \caption{\small (Color online)  Boltzmann conductivity, Seebeck coefficient, and electron thermal coefficient for Silicon calculated with BoltzWann and with our approach.}
    \label{conductivity}
  \end{center}
\end{figure}

Generalizing  Eq. (\ref{boltzmann.conductivity.equation.generalized}) it is also possible to define analogue  expressions for the Seebeck-coefficient $S$ and the electron contribution to thermal conductivity $\kappa_{el}$. 
Following the notation of Ref. \onlinecite{Mecholsky:2014jv}, we  introduce the { generating tensors} ${\mathcal L}_{\alpha}$ ($\alpha = 0, 1, 2$):
\begin{equation} \label{generating.tensors}
 \mathcal L_{\alpha} = \frac{1}{4\pi^3} \int \tau \sum_n  {\bf v}_n({\bf k})\left[{\bf v}_n({\bf k}) \cdot \hat{{\bf e}}\right] \left( -\frac{\partial f_0}{\partial \epsilon}\right)\left[\epsilon_n-\mu\right]^{\alpha}d {\bf k},
\end{equation}
where $\hat{{\bf e}}$ is the direction of the external electric field and $\mu$ is the chemical potential. The coefficients $\sigma$, $S$ and $\kappa_{el}$ can be expressed as follows:
\begin{eqnarray}
\sigma &=& e^2\mathcal L_0 \\ \nonumber
S     &=& -\frac{1}{T e} \left[\mathcal L_0\right]^{-1} \cdot \mathcal L_1 \\ \nonumber
\kappa_{el} &=& \frac{1}{T} \left(\mathcal L_2 - \mathcal L_1 \cdot \left[\mathcal L_0\right]^{-1} \cdot \mathcal L_1\right),
\end{eqnarray}
where $T$ is the temperature. 
From Eqs (\ref{boltzmann.conductivity.equation.generalized}-\ref{generating.tensors}) it is evident that the evaluation of the transport properties requires an accurate integration
over a fine grid of k-point in the BZ, especially for highly dispersive bands as in metal systems. This becomes a trivial task using the TB representation from the PAO projections and Eq. (\ref{Hamiltonian.gradient}).
As a validation of this approach we have calculated the transport coefficients ($\sigma$, $S$, $\kappa_{el}$) of Silicon and compared with the results of the code BoltzWann, where the interpolation of the real space Hamiltonian is done in using Maximally Localized Wannier Functions as basis functions.\cite{Pizzi:2014ee} The results are summarized in Fig. \ref{conductivity} and show excellent agreement between the two approaches. 

\subsection{Ballistic transport}\label{ballistic}

Calculations of the ballistic electrical conductance {\it \`a la} Landauer are naturally built on a local representation of the electronic structure like the one provided by TB hamiltonians. Our procedure reduces the problem of calculating electron transport \cite{Nardelli:1999p58,Calzolari:2004p39} to a computationally inexpensive post-processing maintaining the predictive power and the accuracy of first principles methods. Briefly, using the Landauer approach the conductance is determined via the transmission function that can be
written as: \cite{Fisher:1981gm,Nardelli:1999p58}

\[
\mathcal{T}_{el} = {\rm Tr}(\Gamma_L G_C^r \Gamma_R G_C^a),
\]
where $G_C^{\{r,a\}}$ are the retarded and advanced Green's functions
of the conductor, respectively, and $\Gamma_{\{L,R\}}$ are functions
that describe the coupling of the conductor to the leads. The Green's
function for the whole system can be explicitly written
as:\cite{Datta:1997tk}

\begin{equation}
G_C = (\epsilon -H_C -\Sigma_L -\Sigma_R)^{-1}
\label{gconduct}
\end{equation}

where $\Sigma_{L}$ and $\Sigma_R$ are the self-energy terms due to the
semi-infinite leads.

Once the self-energy functions are known, the coupling functions
$\Gamma_{\{L,R\}}$ can be easily obtained as\cite{Datta:1997tk}

\[
\Gamma_{\{L,R\}} = {\rm i}[\Sigma_{\{L,R\}}^r - \Sigma_{\{L,R\}}^a].
\]

The expression of the self-energies can be deduced along the lines of
Ref. \onlinecite{Nardelli:1999p58} using the formalism of principal layers in the
framework of the surface Green's function matching
theory. We obtain:

\begin{equation}
\begin{array}{ccl}
\Sigma_L & = & H_{LC}^\dagger (\epsilon -H_{00}^L-(H_{01}^L)^\dagger
\overline T_L)^{-1} H_{LC}\\ \Sigma_R & = & H_{CR} (\epsilon
-H_{00}^R-H_{01}^R T_R)^{-1} H_{CR}^\dagger,\\
\end{array}
\end{equation}

where $H_{nm}^{L,R}$ are the matrix elements of the Hamiltonian
between the layer orbitals of the left and right leads respectively,
and $T_{L,R}$ and $\overline{T}_{L,R}$ are the appropriate transfer
matrices. The latter are easily computed from the Hamiltonian matrix
elements via an iterative procedure.\cite{Nardelli:1999p58} This approach has been extensively validated and it is standard procedure in many electronic structure software packages such as WanT,\cite{Calzolari:2004p39} Wannier90,\cite{Mostofi:2014hw} and Smeagol.\cite{Rocha:2006fk}

\subsection{Dielectric function} 

The optical properties of a material are generally described in semiclassical  linear response theory by the dielectric tensor $\epsilon(\omega, {\bf q})$ that is a complex function describing the optical-response of the material in the presence of an external  electromagnetic field at a given
frequency $\omega$ and momentum ${\bf q}.$\cite{Parravicini:2000ud,Onida:2002zz}
Quantities such as the refraction index and the absorption spectrum are easily derived from the real and imaginary part of the dielectric tensor.
In the limit of long wavelength (i.e. negligible momentum transfer ${\bf q}$), 
the optical properties of the material depends only on the frequency of the field.
The dielectric tensor can then be expressed in terms of the dielectric
susceptibility $\chi_{ij}(\omega)$:

\begin{equation} \label{dielectric.function}
\epsilon_{ij}(\omega) = 1 + 4 \pi \chi_{ij}(\omega).
\end{equation}

Following the seminal work by M. Graf and P. Vogl,\cite{Graf:1995tl}  the imaginary part of $\chi(\omega)$ in the single particle approximation can be written as:

\begin{eqnarray} \label{imaginary.dielectric.susceptibility}
\operatorname{Im} \chi_{ij}(\omega) &=& \frac{e^2 \pi}{\omega^2 \hbar m_0^2 \Omega} \sum\limits_{n,m,{\bf k}} \left[f_n({\bf k}) - f_m({\bf k}) \right] \cdot\\ \nonumber
                      & & p_{nm}^i({\bf k})p_{mn}^j({\bf k}) \delta(\omega - \omega_{mn}({\bf k})).
\end{eqnarray}

where $m_0$ is the bare electron mass, $\Omega$ the unit cell volume, $m,n$ the band indices, $f_{\ell}({\bf k})$ the Fermi-Dirac
distribution evaluated on the band with index ${\ell}$ at energy $E_{\ell}({\bf k})$,  $p_{nm}^i({\bf k})$ are the matrix elements
of the momentum operator calculated over the states (both occupied and empty) with indices $m$ and $n$ and $\hbar \omega_{mn} = E_m({\bf k})-E_n({\bf k})$ is the energy of the optical transition.
The real part of the dielectric susceptibility can then be expressed as the Kramers-Kronig transformation of the imaginary part

\begin{equation} \label{KK.imaginary.dielectric.susceptibility}
\operatorname{Re}\chi(\omega) = \frac{2}{\pi} \int\limits_{0}^{\infty}  z \frac{\operatorname{Im}\chi(\omega)}{z-\omega}dz.
\end{equation}

Eq.~(\ref{imaginary.dielectric.susceptibility}) implicitly contains both intra- ($n=m$)  and inter-band ($n \ne m$)
transitions. Inter-band transitions  are associated to the usual optical absorption processes in the UV-visible range,
while intra-band transitions are relevant in
the low frequency regime.
In the latter case, Eq.~(\ref{imaginary.dielectric.susceptibility}) naturally discriminates the different trends of quasi-static dielectric function ($\epsilon (\omega \rightarrow 0)$)
for insulating (finite behavior) and metallic (diverging behavior) systems.
In order to separate the intra- and inter-band contributions, we rewrite the expression $[f_n({\bf k}) - f_m({\bf k})]$  that appears in Eq.~(\ref{imaginary.dielectric.susceptibility}).
The presence of delta function selects the energy of the transitions  $E_n({\bf k+q}) = E_n({\bf k}) + \hbar \omega$ and, consequently, it fixes  the argument of the Fermi-Dirac distributions. Thus for $n = m$ we obtain:

\begin{eqnarray} \label{approximation.for.fermi.difference}
f_n({\bf k}) - f_n({\bf k + q}) &=& \\ \nonumber
f(E_n({\bf k})) -f(E_n({\bf k})+ \hbar \omega) &\approx& \\ \nonumber
 - \hbar \omega \frac{\partial f(E)}{\partial E}|_{E=E_n(\bf k)}.
\end{eqnarray}

In the case of undoped semiconductors and insulators the intra-band transitions do not contribute to Eq.~(\ref{imaginary.dielectric.susceptibility}),
the imaginary part of the dielectric function vanishes, while the real part tends to static dielectric constant.
For metallic systems the presence of partially occupied bands at the Fermi energy 
makes the derivative in Eq.~(\ref{approximation.for.fermi.difference})  converging to a finite quantity, while the term $\frac{1}{\omega^2}$  in Eq.~(\ref{imaginary.dielectric.susceptibility}) diverges. The divergency of the dielectric function gives rise to the well-known Drude-like {\em dc}-conductivity of metals, where electrons close to Fermi-level can undergo electron transitions 
with negligible momentum transfer in the quasi-static regime ($\omega \rightarrow 0$). 

The evaluation of the momentum matrix elements, $p_{nm}^i({\bf k})$, is a computational bottleneck since it requires an integration over all the pairs of occupied and empty electronic states across the whole BZ.
If the single particle wavefunctions are expanded in 
large basis sets (e.g. planewaves), as in standard solid state implementations,  this integration rapidly becomes a computational challenge. Once again, the momentum can be efficiently evaluated using the TB Hamiltonian projected on the PAOs, transforming the calculation of the frequency dependent dielectric function into a computationally trivial post-processing of the first principles calculation.

\begin{figure}[h!]
\begin{center}
    \includegraphics[width=0.95\columnwidth]{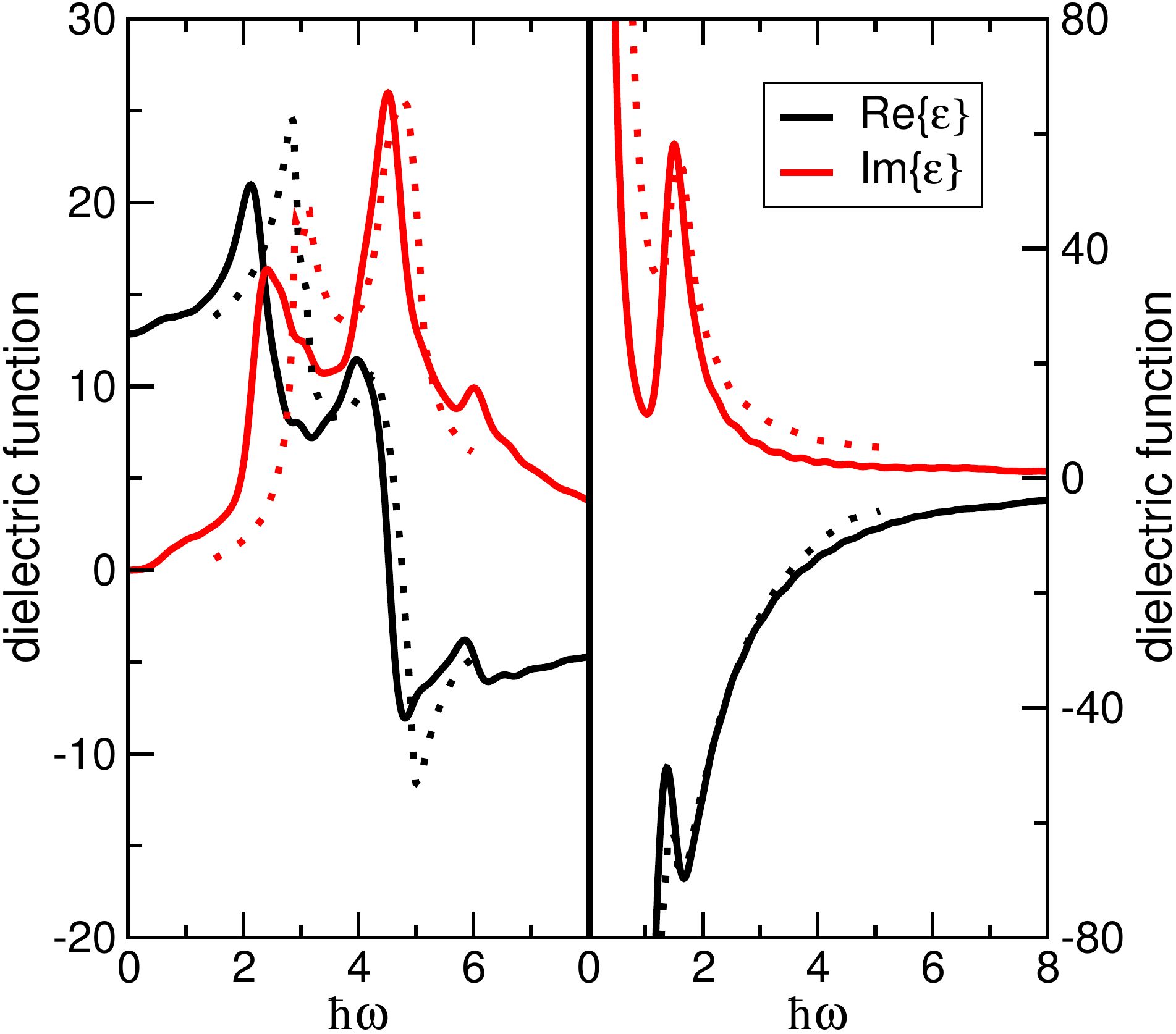}
    \vspace{5mm}
    \caption{\small (Color online)  Real and Imaginary part of the dielectric function for GaAs (left panel) and Al (right panel). Results are validated against experimental data: Ref. \onlinecite{Ehrenreich:1963cq} for Al and Ref. \onlinecite{Aspnes:1983fy} for GaAs.}
    \label{optical}
  \end{center}
\end{figure}

We have validated the method for calculating the dielectric function against well known experimental results for GaAs and Al. Results are summarized in Fig. \ref{optical} and the comparison  confirms the validity of our approach.



\subsection{The ACBN0 functional} \label{acbn0}

The knowledge of first principles based localized orbital representation  of the electronic structure combined with analytical expressions based on Gaussian basis set provides an efficient strategy for the fast computation of two-electron integrals for solid-state applications and the development of local exchange functionals (LEX). This is a critical advantage when dealing with the plethora of novel materials that are characterized by strong electron localization
and correlation and vigorously sought for their rich physical and chemical properties. For these materials the LDA+$U$ method, introduced by Liechtenstein and Anisimov, \cite{Anisimov1997,Liechtenstein1995} is the most practical choice to compensate for the simplified, nearly-homogeneous-electron-gas treatment of the electron density by LDA.
The success of LDA(GGA)+$U$ confirms that preserving the information of orbital localization from being averaged out is prevalent to the correct prediction of the electronic structure in compounds such as transition metal oxides.\cite{prbpriya:prep}

Our projection methodology allows the direct computation of two-electron integrals and, when combined with a density-matrix-based approach, the direct and self-consistent evaluation of the on-site Coulomb $U$ and exchange $J$
parameters needed in the treatment of correlated solid materials. This is at the core of the definition of the ACBN0 functional, recently introduced by some of us.\cite{Agapito:2015iz}  ACBN0 satisfies the rather ambitious criteria outlined by Pickett {\it et al.}\cite{Pickett:1998tu} in one of the first seminal articles on LDA+$U$. 
Due to the projection on AOs and the accurate TB representation,  the evaluation of  the \textit{U} and \textit{J} for atoms in different chemical environments or close to topological defects (surfaces, interfaces, impurities, etc.) or for closed-shell atoms (like Zn) becomes trivial, thus overcoming the limitations of traditional linear response techniques.\cite{reviewLDAU}
Results so far are striking: comparisons with available experimental and theoretical data show that the proposed computation of the on-site Coulomb and exchange parameters is a sound and high-throughput alternative to higher levels of theory such as hybrid functionals and the $GW$ approach that systematically yields results with outstanding accuracy.\cite{Agapito:2015iz,Gopal:2015bf}
ACBN0 only demands computational resources comparable to a regular (LDA)PBE calculation. 

\section{Computational details.} \label{comp_details}

Ground state properties are obtained with density functional theory as implemented in the {\small Quantum ESPRESSO} package.\cite{Giannozzi:2009p1507}  The starting electronic structure has been determined using ACBN0 and compared, when appropriate, with traditional PBE. 
Ionic potentials are described by pseudopotentials with an extended basis set of pseudo-atomic orbitals for an improved TB mapping of the conduction bands.\cite{Agapito:2016en} 
The workflow to perform all steps of the approach is part of the {\sf AFLOW$\mathlarger{\mathlarger{\mathlarger{{\pi}}}}$} software infrastructure.\cite{AFLOWpi:prep} {\sf AFLOW$\mathlarger{\mathlarger{\mathlarger{{\pi}}}}$}
has been designed for high-throughput first principles calculations and exploits the projections on PAO and the TB representation for the calculation of the band structure, density of states, transport coefficients, and frequency dependent dielectric constant. {\sf AFLOW$\mathlarger{\mathlarger{\mathlarger{{\pi}}}}$}
 is integrated with AFLOW.\cite{Curtarolo2012218} 

\section{Results} \label{results}
Our TB representation enables studies that involve extremely fine sampling of the full $E_n({\bf k})$ electronic structure and brings within reach nanostructured systems with hundreds of atoms without compromising first principles accuracy.
As examples to validate our methodology we will analyze with unprecedented details the conduction manifold of CoSb$_3$-based thermoelectric materials (Sec. \ref{cosb3}) and a core-shell nanowire with 588 atoms (Sec. \ref{coreshell}).

\subsection{CoSb$_3$}\label{cosb3}
Materials with the skutterudite structure are of great interest for their performance as thermoelectrics (TEs) since they well represent the paradigmatic case of a phonon-glass electron-crystal.\cite{Singh2001125,NolasBook,doi.org/10.1038/nmat2090,ANIE:ANIE200900598,doi:10.1179/095066003225010182}
They exhibit low thermal conductivity but also excellent electronic properties
and have been studied extensively both theoretically\cite{Singh:prb1994, PhysRevB.81.045204} and experimentally.\cite{doi:10.1146/annurev.matsci.29.1.89,Uher2001139}
The thermoeletric performances are characterized by the so-called figure of merit, $ZT=S^2\sigma T/\kappa$, where $S$ is the Seebeck coefficient, $\sigma$ is the electrical conductivity, $\kappa$ is the thermal conductivity, and 
$T$ is the temperature at which the device operates. 
The prototypical $n$-type compound within the skutterudites family is CoSb$_3$.\cite{doi:10.1146/annurev.matsci.29.1.89,Caillat:japphys1996, Tang:jmateriom2015}
Indeed, the figure of merit of  R-filled CoSb$_3$ with R = Na, Yb, In, Ba, Ce 
exceeds one in the temperature windows of practical interest.
Typical descriptors for enhanced TE electronic properties include reasonable large effective masses which favor large Seebeck coefficients and multi-valley character of the active bands to optimize the conductivity.\cite{Singh2001125,doi:/content/aip/journal/apl/74/24/10.1063/1.124220,doi:10.1021/acs.chemmater.5b04365}  Doping and filling the semiconducting CoSb$_3$ with donor elements activates  the bands at the bottom of the conduction manifold whose features critically contribute to optimize the value of electronic transport coefficients. The enhanced performance of RCoSb$_3$ can be rationalized with a detailed analysis of the full $E_n({\bf k})$ for all the bands in the proximity of the Fermi level. Various experimental and theoretical investigations have been performed in order to highlight the mechanisms responsible for its good thermoelectrical performance and the physics of the system is still under debate. In this paper we extend previous studies on the conduction band of doped CoSb$_3$ (Ref. \onlinecite{Tang:nmat2015}) by taking advantage the improved band structure provided by the ACBN0 functional discussed in Sec. \ref{acbn0} and the Boltzmann transport capabilities presented Sec. \ref{boltzmann}.

\begin{figure}[h!]
\begin{center}
    \includegraphics[width=0.99\columnwidth]{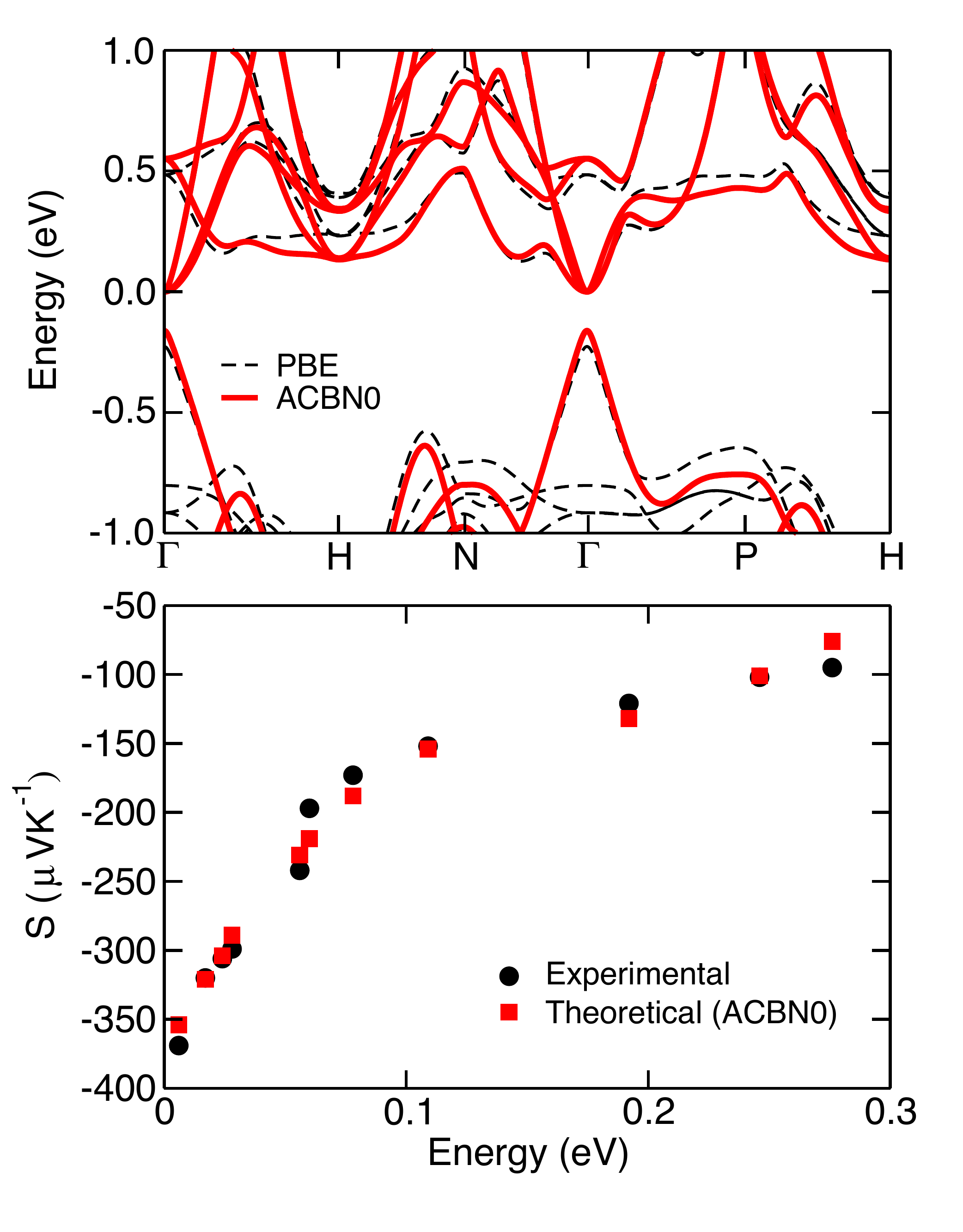}
    \vspace{1mm}
    \caption{\small (Color online)  (Top panel) Electronic structure of CoSb$_3$. The black-dashed and the red-solid lines refers to PBE\cite{Tang:nmat2015} and ACBN0 calculations. (Botton panel) Seebeck coefficient of CoSb$_3$ measured experimentally (black filled circles)\cite{Tang:nmat2015} and computed theoretically (red filled squares) with a two-effective-masses model. The energy is measured from the bottom of the conduction band.}
    \label{fig:cosb3}
  \end{center}
\end{figure}

We have computed the starting electronic structure of CoSb$_3$ (prototype: A3B\_cI32\_204\_g\_c in Ref. \onlinecite{Mehl:2016wf})  using a {\bf k}-point grid of ($9 \times 9 \times 9$).\cite{Tang:nmat2015} The Boltzmann transport coefficients have been then calculated using a much finer grid with ($100 \times 100 \times 100$) {\bf k}-points.
In Fig. \ref{fig:cosb3}-(a) we compare the band structure of the system calculated with PBE and ACBN0. Qualitatively, the bands compare well with previous calculations\cite{Singh:prb1994}, however, in our ACBN0 calculations the energy gap between the occupied and the unoccupied manifold decreases from the PBE value of 0.23 eV to 0.16 eV. Both values are consistent with the experimental findings that range from 0.05 eV to 0.22 eV,\cite{Pei:advmat2012} but the reduction of the energy gap may be important to investigate effects associated with decreased performance due to bipolar transport.
The linearity of the dispersion\cite{Singh:prb1994} at the top of the valence band ($p$-derived band near $\Gamma$) slightly increases and the occupied Co-$d$ manifold  moves to lower energy. This is consistent with the fact that the Sb Hubbard correction (U$^{Sb}$ = 0.648 eV) is small whereas the value of U for the Co cation (U$^{Co}$ = 4.375 eV) leads to a larger variation. This result is of particular importance when chemical substitution is used to reduce the energy separation between dispersive and flat bands in order to optimize the $p$-type transport coefficients.
The description of the conduction manifold also improves. In Fig. \ref{fig:cosb3}-(upper panel) the valleys at H and along the $\Gamma$-N direction in the BZ become quasi-degenerate ({$\Delta E_{ACBN0} = 10$ meV}) and flatten. It is remarkable that this quasi-degeneracy is not captured by standard PBE calculations ($\Delta E_{PBE} = 100$ meV).\cite{Tang:nmat2015} Phenomenologically these effects contribute to increase the number and the effective mass of charge carriers with positive consequences on the conductivity and Seebeck coefficient.

Within the rigid band approximation, we computed the Seebeck coefficient at values of the chemical potential corresponding to the experimental electron density\cite{Tang:nmat2015} and we compare our theoretical prediction with the experimental results as shown in Fig. \ref{fig:cosb3}-(lower panel). 
In order to obtain the value of $\mu$ corresponding to a given experimental electron density $n_e$, we have used the free-electron 3D gas relationship $\mu = \frac{\hbar^2}{2 m^*}(3\pi^2n_e)^{2/3}$ where we have employed two effective masses calculated from the electronic structure at $\Gamma$ and at H assuming a parabolic dispersion.
Results are summarized in Fig. \ref{fig:cosb3}-(b) where we observe an excellent agreement between the calculated and experimental value of the Seebeck coefficient S. 

\subsection{Core-shell nanowire}\label{coreshell}

Quasi-one-dimensional, vertically aligned nanowires can be exploited to construct three-dimensional architectures with demonstrated advantages over conventional planar devices.  Nanowires form building blocks for compact ultrafast electronics and optoelectronic devices (e.g. solar cells \cite{garnett10,shu13}, photodetectors\cite{soci10}, nanoscale lasers,\cite{duan03}  and light emitting diodes\cite{bao06, sarwar15}). 
One-dimensional (1D) component-modulated materials, such as coaxial core-shell heterojunctions  offer the benefit of designing and fabricating nanodevices without further assembling and provide unique and tunable properties.\cite{lauhon02,boland15,Saha:2013fy} 

\begin{figure}[h!]
\begin{center}
    \includegraphics[width=0.99\columnwidth]{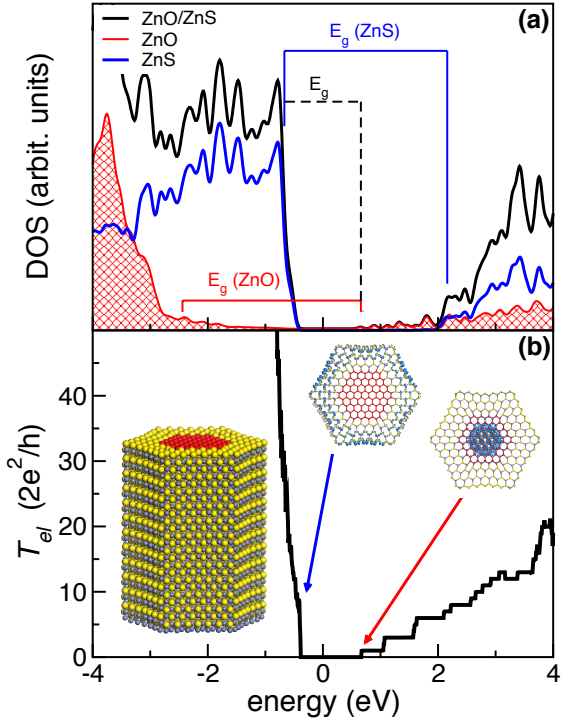}
    \vspace{5mm}
    \caption{\small (Color online)  (a) Total and material-projected density of states plot; (b) Quantum conductance. Left inner panel: geometry of the CCS nanowire. Right inner panels: eigenchannels of the trasmission amplitude corresponding to the top of the valence and bottom of conduction bands. Zero energy reference is aligned to the Fermi level of the system.}
    \label{wire}
  \end{center}
\end{figure}

To highlight the capability of our methodology in evaluating the optical and transport properties of nanoscale systems, we have chosen as a prototypical example a coaxial core-shell (CCS) nanowire of ZnO and ZnS (see Fig. \ref{wire}(b), left inner panel) that has been recently synthesized\cite{Brayek:2014bt,Huang:2013kj,Nam:2011kl,Shuai:2011jb,Zhu:2008bi}.  

ZnO has a wide band gap energy of 3.37 eV and it possesses unique optical and electronic properties that make it a promising candidate for UV lasers and detectors working in the 320-400 nm wavelength range. Additionally, it is transparent to visible light and can be made highly conductive by doping.\cite{wang04}
ZnS is also a wide band gap semiconductor, with bandgap energy of 3.66 eV widely exploited for optoelectronic devices and sensors. The electrical and the optical characterization of ZnS coated ZnO nanowires have been studied extensively both experimentally\cite{li06,wang10,meng10,Fang:2015kb,Fang:2016ep,Jeong:2014fr} and theoretically, although on much smaller diameter wires.\cite{Schrier:2007ds}

In this study, the core-shell nanowire is simulated using the ACBN0 functional in a large cell ($50.0 \times 50.0 \times 5.3$) A$^3$ with  588 atoms (i.e. ~5300 electrons). The core is made of ZnO and it has an internal radius $r_c$=1.1 nm, that is large enough to correctly reproduce a realistic ZnO wire\cite{PhysRevB.80.201304}. The shell is made of ZnS, the total radius of the heterostructure is $r_s$=1.9 nm, in agreement with  experimental samples\cite{li06}. The wire is aligned along the polar c-axis of the wurtzite ZnO crystal. Due to the huge dimension of the system, we include two bilayers of ZnS-ZnO wurtzite material along the wire direction, \textit{i.e.} the minimum to obtain a periodic wire. The CCS structure has a hexagonal symmetry and exposes only non-polar (10$\bar{1}$0) faces.
The geometry is fully relaxed until forces on all atoms are lower than 0.03 eV/\AA. In the optimized structure the inner ZnO core almost maintains its ideal geometry, while the external ZnS shell undergoes to remarkable distortion due to the relaxation
of the mismatch at the interface. Nonetheless the outermost layer exhibits a buckled dimer arrangement, typical of the ZnS(10$\bar{1}$0) surface (see inset of Fig. \ref{wire}).

The resulting ACBN0 electronic structure is summarized in Fig.  \ref{wire}a, where we plot the total (black line) and ZnO (red shaded area) and ZnS (blue thin line) projected density of states (DOS). Albeit ZnO and ZnS have similar band gap, the different ionization potential causes the formation of a staggered type-II band-alignment at the interface, with the top of ZnS valence band lying in the pristine gap of ZnO material. The ZnO core has a band gap E$_g\sim 3.1$ eV very similar to corresponding bulk, while the outer ZnS layer has a band gap  E$_g\sim2.8$ eV. This gap reduction derives from the strong atomic deformation, which makes the final structure sensitively different from the ground state bulk one. The total band gap of the CCS nanowire is E$_g\sim 1.2 $ eV that lies in the near-IR range. 

By using the approach described in Sec. \ref{ballistic}, we calculated the coherent electron transport along the nanowire. In the minimal TB-representation the solution of the Landauer problem reduces to matrix operations between (2940$\times$2940) on-site and hopping hamiltonians,\textit{i.e.}  much smaller than the corresponding plane wave ones. This makes an otherwise unsolvable problem computationally feasible. 

\begin{figure}[h!]
\begin{center}
    \includegraphics[width=0.95\columnwidth]{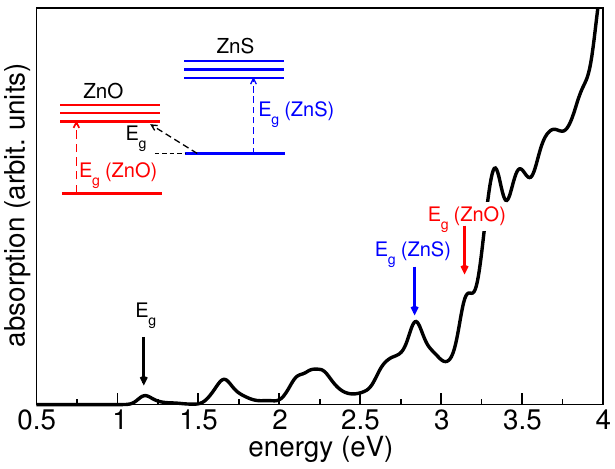}
    \vspace{5mm}
    \caption{\small (Color online)  Imaginary part of dielectric function of ZnS/ZnO CCS nanowire. Inset highlights band-alignment scheme and lowest energy vertical transitions.}
    \label{wire_opt}
  \end{center}
\end{figure}

For the nanowire we are considering here, the scattering contributions with the boundaries, due to extremely high surface-to-volume ratio, are the predominant effects that control the electron transport.\cite{rcsAvd} Thus, the coherent regime is a good  first approximation for the description of transport, at least at low temperature. 
The result for the CCS nanowire is shown in Figure \ref{wire}b.
The  quantum transmittance $\mathcal{T}_{el}$ is proportional to the number of transmitting channels available for electron mobility, which are equal to the number of conducting bands at the same energy. 
Within the scattering theory framework, the transmittance function $\mathcal{T}_{el}$ can be related to the transmittance amplitude $\mathrm{t}$ through the relation $\mathcal{T}_{el}=Tr[\mathrm{t}^{\dag}\mathrm{t}]$. The eigenvectors 
of matrix $\mathrm{t}$ are called {\em eigenchannels} and 
are defined as the linear combinations of the incoming modes in a lead that do not mix upon reflection on the scattering region. Thus, the spatial representation of the eigenchannels visually displays
the path  travelled by charge carriers in their flow through the nanowire. The eigenchannels corresponding to the top of the valence and bottom of conduction bands are shown as inset in panel b: hole carriers flow in the external ZnS crown, while electrons move in the internal ZnO core, confirming the intrinsic charge separation observed experimentally.\cite{Jeong:2014fr}

Simulated optical properties are summarized in Fig \ref{wire_opt}. The peaks in the imaginary part of the dielectric function $\epsilon$ correspond to single particle valence-to-conduction transitions. As expected, it is easy to recognize  the valence-to-conduction absorption edge in the external ZnS shell (2.8 eV, blue arrow) and in the inner ZnO (3.1 eV, red arrow), 
which correspond to the E$_g$ values of each material, as discussed above. However, other lower-energy transitions are present in the range 
1.2-2.8 eV, which correspond to inter-material ZnO-to-ZnS transition as depicted in the inset of the figure. Although the frontiers orbitals are mostly localized in the core (electrons) and in the shell (hole), the overlap and the symmetry of the wave functions
give low but not negligible oscillator strength to the first four transition. Except for excitonic effects (not included at this level of theory), this agrees well with dramatic red-shift of the absorption edge (i.e. from UV - to near-IR) observed experimentally
\cite{Fang:2015kb,Fang:2016ep}, also in agreement with previous theoretical calculations  on smaller CCS wires\cite{Schrier:2007ds}. The facile photocharge injection and the intrinsic carrier separation make this system a very promising photo-conductor candidate for optoelectronic and photovoltaic devices.

\section{Conclusions} \label{conclusions}

We have extended the PAO projection technique to allow for the calculation of electronic transport and optical properties of materials with extreme accuracy and negligible computational cost. The exact tight binding representation of the first principles electronic structure allows on one hand to produce extremely dense band interpolations, essential requirement for the evaluation of Boltzmann transport or optical properties, and on the other, the local Green's function representation that is at the foundation of quantum conductance calculations.
We have demonstrated the potentiality of the method by studying  the multi-valley band structure of CoSb$_3$ and a large ZnO-ZnS core-shell nanowire. 

\acknowledgments
We want to thank the Texas Advanced Computing Center (TACC) at the University of Texas Austin and CINECA for providing computing facilities; the funding provided by DOD-ONR (N00014-13-1-0635, N00014-15-1-2266 and N00014-14-1-0526); the support from the University of Modena and Reggio Emilia through the grant ``Nano- and emerging materials and systems for sustainable technologies''; and the European Union Seventh Framework Program for the grant agreement 265073 (ITN-Nanowiring). The authors also acknowledge the Duke University Center for Materials Genomics and the CRAY Corporation for computational assistance.


\end{document}